\documentclass[longbibliography, prl, amsmath, amssymb,superscriptaddress,fleqn]{revtex4-2}
\usepackage{amsfonts}
\usepackage{graphicx}
\usepackage[colorlinks,linkcolor=blue,citecolor=blue,urlcolor=blue]{hyperref}
\usepackage{float}
\usepackage[normalem]{ulem}

\begin{document}

\title{Limit Cycles and Chaos Induced by a Nonlinearity with Memory}

\author{K. J. H. Peters}\email{k.peters@amolf.nl}
\affiliation {Center for Nanophotonics, AMOLF, Science Park 104, 1098 XG Amsterdam, The Netherlands}

\author{S. R. K. Rodriguez}  \email{s.rodriguez@amolf.nl}
\affiliation {Center for Nanophotonics, AMOLF, Science Park 104, 1098 XG Amsterdam, The Netherlands}

\begin{abstract}
    Inspired by the observation of a distributed time delay in the nonlinear response of an optical resonator, we investigate the effects of a similar delay on a noise-driven mechanical oscillator. For a delay time that is commensurate with the inverse dissipation rate,  we find stable limit cycles. For longer delays, we discover a regime of chaotic dynamics associated with a double scroll attractor. We also analyze the effects of time delay on the spectrum and oscillation amplitude of the oscillator. Our results point to new opportunities for nonlinear energy harvesting, provided that a nonlinearity with distributed time delay can be implemented in mechanical systems.  
\end{abstract} 
\date{\today}
\maketitle

Despite what Newton's laws suggest, physical systems do not respond instantaneously. Real systems generally have a non-instantaneous response, which can be mathematically described by a time-delayed term in the equation of motion representing them~\cite{popovych2011delay,kuang2012delay,liu2016dynamics,zhang2019recent,keane2017climate,Terrien18,Otto19,Hart19}.
Already a simple constant time delay can result in complex behavior, such as delay-induced bifurcations~\cite{xu2004delay,shayer2000stability,liao2001local} and chaos~\cite{an1983existence,safonov2002delay}. The more general classes of distributed time delays~\cite{wang2004global,sun2006inducing,jeevarathinam2015vibrational,cantisan2020delay,coccolo2021delay}, time-varying delays~\cite{louisell2001delay,botmart2012synchronization,ardjouni2012existence,muller2018laminar,muller2019resonant} and state-dependent delays~\cite{insperger2007state,keane2019effect} can can also lead to a rich phenomenology.   Beyond their fundamental relevance, time-delayed systems are also relevant to many applications in computation and machine learning~\cite{ortin2015unified,hart2019delayed,penkovsky2019coupled,Barbay20,Harkhoe20,brunner2021nonlinear,banerjee2021machine}, sensing~\cite{grigor1985laser,zou2015optoelectronic,yao2017optoelectronic}, and chaos-based communication~\cite{larger2004ikeda,li2004chaotic,jiang2011chaos}.  While a number of systems with nonlinear time delay have garnered strong interest~\cite{mackey1977oscillation,ikeda1979multiple,lepri1994high,muller2018laminar,Hart19PRL},  most efforts in the field have focused on systems with time delay in their linear response.

Thermo-optical nonlinear cavities offer a convenient platform for probing an interesting type of nonlinear $distributed$ time delay~\cite{Yacomotti06,Alaeian17,geng2020universal,peters2021extremely}. Optical experiments revealed that a  nonlinearity with distributed time delay leads to a universal scaling law for hysteresis phenomena~\cite{geng2020universal}, and an enlarged bandwidth for noise-assisted amplification of periodic signals~\cite{peters2021extremely}.  Motivated by these recent findings of fundamental and practical relevance, here we explore the physics of a noise-driven oscillator with variable distributed time delay in its nonlinear response. We demonstrate the emergence of stable limit cycles when the delay time is commensurate with the inverse dissipation rate of the oscillator. For larger delay times, we discover a chaotic regime associated with a double scroll attractor. Our results have implications for nonlinear vibration energy harvesters~\cite{cottone2009nonlinear}, which can be improved by a nonlinear response with distributed time delay.

\begin{figure}[!t]
	\includegraphics[width=\textwidth]{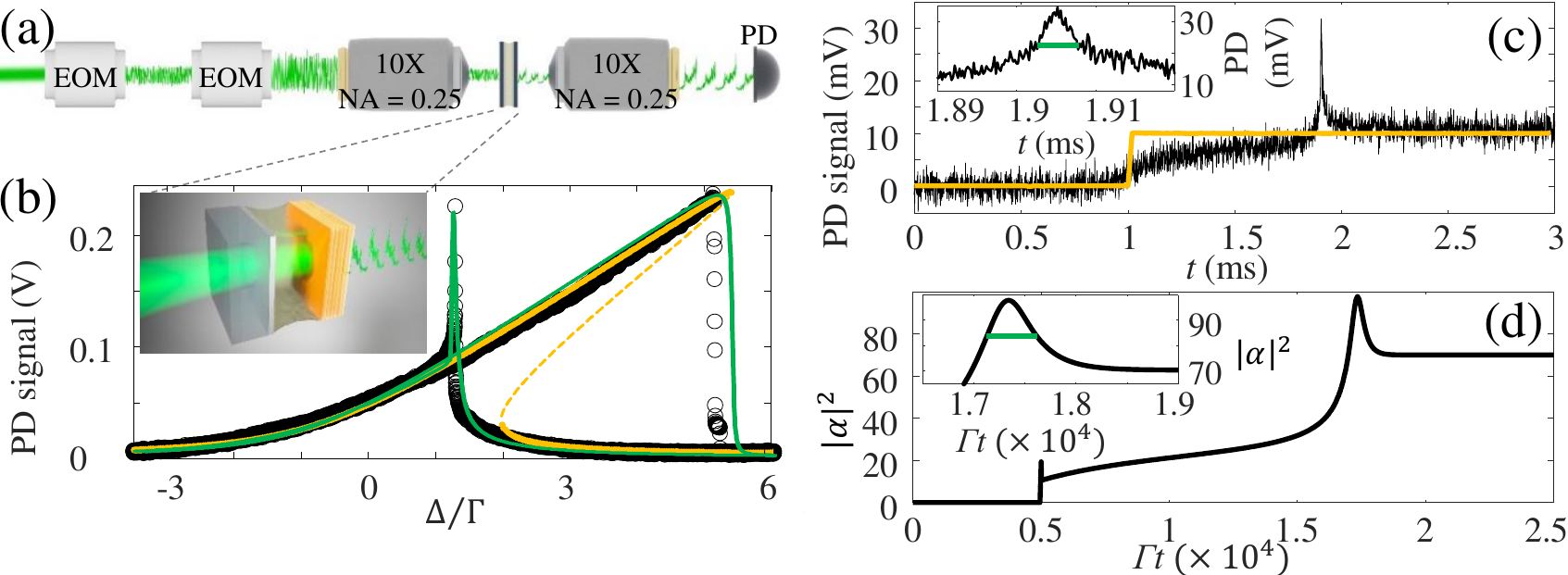}
	\caption{(a) Schematic of our optical setup. Two electro-optical modulators (EOM) add noise to the laser amplitude and phase. The laser light is then focused on an oil-filled optical microcavity using microscope objectives. The transmitted light is measured using a photodetector (PD). (b) Transmitted intensity while opening and closing the cavity (black circles), averaged over 20 cycles. Yellow solid (dashed) curves show stable (unstable) steady-state solutions of Eq.~\ref{eq:cavity}. Green solid curve shows a dynamical simulation while scanning $\Delta/\Gamma$. Inset: Schematic of an oil-filled optical microcavity. (c) Transmitted intensity (black) when the input laser is modulated by a chopper, creating a step function in the intensity (yellow). Inset: Zoom of overshoot. Green line indicates the full-width half-maximum of the overshoot, which is $5$~$\mu$s. (d) Simulation of Eq.~\ref{eq:cavity} when increasing $F$ from $0$ to $7\sqrt{\Gamma}$ at $\Gamma t=5000$. Inset: Zoom of the overshoot. Green line indicates width of the overshoot, which is $\tau/2$.}\label{fig:1}
\end{figure}

Let us first demonstrate, experimentally, the existence of a distributed time delay in the thermo-optical nonlinear response of an oil-filled optical microcavity. Figure~\ref{fig:1}(a) illustrates our experimental setup: a tunable Fabry-P\'erot cavity filled with macadamia oil and driven by a 532~nm continuous wave laser. The cavity [Fig.~\ref{fig:1}(b) inset] is made by a planar and a concave mirror. The planar mirror comprises a 60~nm silver layer on a glass substrate. The concave mirror has a diameter of $7$~$\mu$m and  a radius of curvature of $12$~$\mu$m. It is fabricated by milling a glass substrate with a focused ion beam~\cite{Trichet15}, and then coating it with a distributed Bragg reflector (DBR). The DBR has a peak reflectance of $99.9$\% at the center of the stopband, located at $530$~nm. Thanks to micron-scale dimensions of the concave mirror strongly confining the optical modes, we can probe a single mode when scanning the cavity length across a wide ($> 10$ nm)  range.

In a frame rotating at the frequency of the driving laser $\omega$, the light field $\alpha$ in our cavity satisfies:
\begin{equation}\label{eq:cavity}
	i \dot{\alpha}(t) = \left[-\Delta- i\frac{\Gamma}{2} + U \int_0^t ds\,K(t-s)\left(|\alpha(s)|^2-1\right)\right]\alpha(t) + i \sqrt{\kappa_L}F + \frac{D}{\sqrt{2}} \left[\xi_1(t) + i \xi_2(t) \right].
\end{equation}
$\Delta=\omega-\omega_0$ is the laser-cavity detuning, with  $\omega_0$ the cavity resonance frequency. $\Gamma=\gamma+\kappa_\mathrm{L}+\kappa_\mathrm{R}$ is the total dissipation rate, with $\gamma$ the intrinsic loss rate, and $\kappa_\mathrm{L}$ ($\kappa_\mathrm{R}$) the input-output coupling rate through the left (right) mirror. $U$ quantifies the strength of the cubic nonlinearity, corresponding to effective photon-photon interactions in optical systems. The memory kernel $K(t)=\mathrm{exp}\left(-t/\tau\right)/\tau$ accounts for the non-instantaneous nonlinear response of our cavity, i.e., the distributed time delay. This time-delayed response arises because the temperature of the oil takes a finite time to relax to a steady state when the laser amplitude $F$ changes.  The term $D\xi(t)=D[\xi_1(t)+i\xi_2(t)]/\sqrt{2}$ represents Gaussian white noise with variance $D^2$ in the laser amplitude and phase. $\xi_i(t)$ each have zero mean [i.e., $\langle \xi_i(t) \rangle = 0$], and are delta-correlated with unit variance [i.e., $\langle \xi_i(t) \xi_i(t+t') \rangle =  \delta(t')$]. Moreover,  $\xi_1(t)$ and $\xi_2(t)$ are mutually uncorrelated. We numerically solve Eq.~\ref{eq:cavity} (and equations ahead) using the xSPDE toolbox~\cite{xSPDE} for Matlab. 

For strong driving (large $F$), the cavity supports optical bistability: two stable steady states with different intra-cavity  intensity $|\alpha|^2$ at a single driving condition. To evidence bistability, we measure the transmitted intensity while scanning the cavity length (and hence $\Delta$) forward and backward. The black curve in Fig.~\ref{fig:1}(b) shows the result  when the laser power is $7.8$ mW at the excitation objective. We observe a large optical hysteresis, and bistability occurs in the $1.5 \lesssim \Delta/\Gamma \lesssim 5$ range.  We also observe a large overshoot around $\Delta/\Gamma=1.5$, which is due to the non-instantaneous thermo-optical nonlinearity of the oil-filled cavity. The solid (dashed) yellow curve in Fig.~\ref{fig:1}(b) shows stable (unstable) steady-state solutions, obtained by setting $\dot{\alpha}=0$ in Eq.~\ref{eq:cavity}. These steady-state calculations reproduce the bistability, but not the overshoot. In contrast, dynamical simulations of Eq.~\ref{eq:cavity}, shown as solid green curves in Fig.~\ref{fig:1}(b), reproduce our experimental observations including the overshoot. The overshoot arises when the duration of the scan is similar to or less than the thermal relaxation time of the oil~\cite{geng2020universal}, which is the case in our experiments. 


Figures~\ref{fig:1}(c,d) further evidence the non-instantaneous nonlinear response of our cavity. The black curve in Fig.~\ref{fig:1}(c) represents the transmitted intensity when modulating the laser power in a step-like fashion, as shown by the yellow curve. Prior to the step, the laser is blocked and the transmission is zero. Immediately after the step, the transmission first increases to a low intensity state. Then, the nonlinearity gradually builds up due to the laser-induced heating of the oil. This results in a slow increase of the transmitted signal. Finally, after the nonlinearity has sufficiently built up, the transmitted intensity displays a large overshoot followed by relaxation to a high intensity steady state. In Fig.~\ref{fig:1}(d) we numerically reproduce our experimental observations using Eq.~\ref{eq:cavity}. From our calculations we find that the full-width at half-maximum of the overshoot, indicated by the green line in the Fig.~\ref{fig:1}(c) inset, is $\tau/2$ regardless of the cavity parameters. Based on this finding, we deduce that the nonlinearity of our oil-filled cavity has a distributed time delay with $\tau=10$~$\mu$s.

The results in Fig.~\ref{fig:1} and in References~\cite{geng2020universal,peters2021extremely} motivate us to explore more generally, beyond the realm of optics, the effects of a nonlinearity with distributed time delay.  In this spirit, we consider a mechanical oscillator with a Duffing-type nonlinearity having distributed time delay. We describe the time delay with  the same kernel function $K(t)$ used to describe our oil-filled cavity.  Thus, our nonlinear oscillator satisfies the following equation of motion:
\begin{equation}\label{eq:Duffing_IDE}
     m\ddot{x}(t)=\left(a+b\int_0^t ds\;K(t-s)x(s)^2\right)x(t)-\gamma \dot{x}(t) + D\xi(t).
\end{equation}
$m$ is the mass of the oscillator and $\gamma$ its dissipation. $a$ and $b$ define the potential $V(x)=-ax^2/2+bx^4/4$ in the limit $\tau\rightarrow 0$. We set $a>0$ and $b>0$, such that $V(x)$ is a double-well potential.


To simulate Eq.~\ref{eq:Duffing_IDE} it is convenient to define the variables $w=b\int_0^t ds\;K(t-s)x(s)^2$ and $v=\dot{x}$. This allows us to write Eq.~\ref{eq:Duffing_IDE} as a set of 3 ordinary differential equations
\begin{equation}\label{eq:Duffing_ODE}
     \begin{split}
         \dot{x}&=v,\\
         m\dot{v}&=\left(a-w\right)x-\gamma v+D\xi(t),\\
         \dot{w}&=\left(bx^2-w\right)/\tau,
     \end{split}
\end{equation}
which we solve numerically using a fourth-order Runge-Kutta algorithm with time increments $\Delta t=\gamma^{-1}/100$.

\begin{figure}[!t]
	\includegraphics[width=\textwidth]{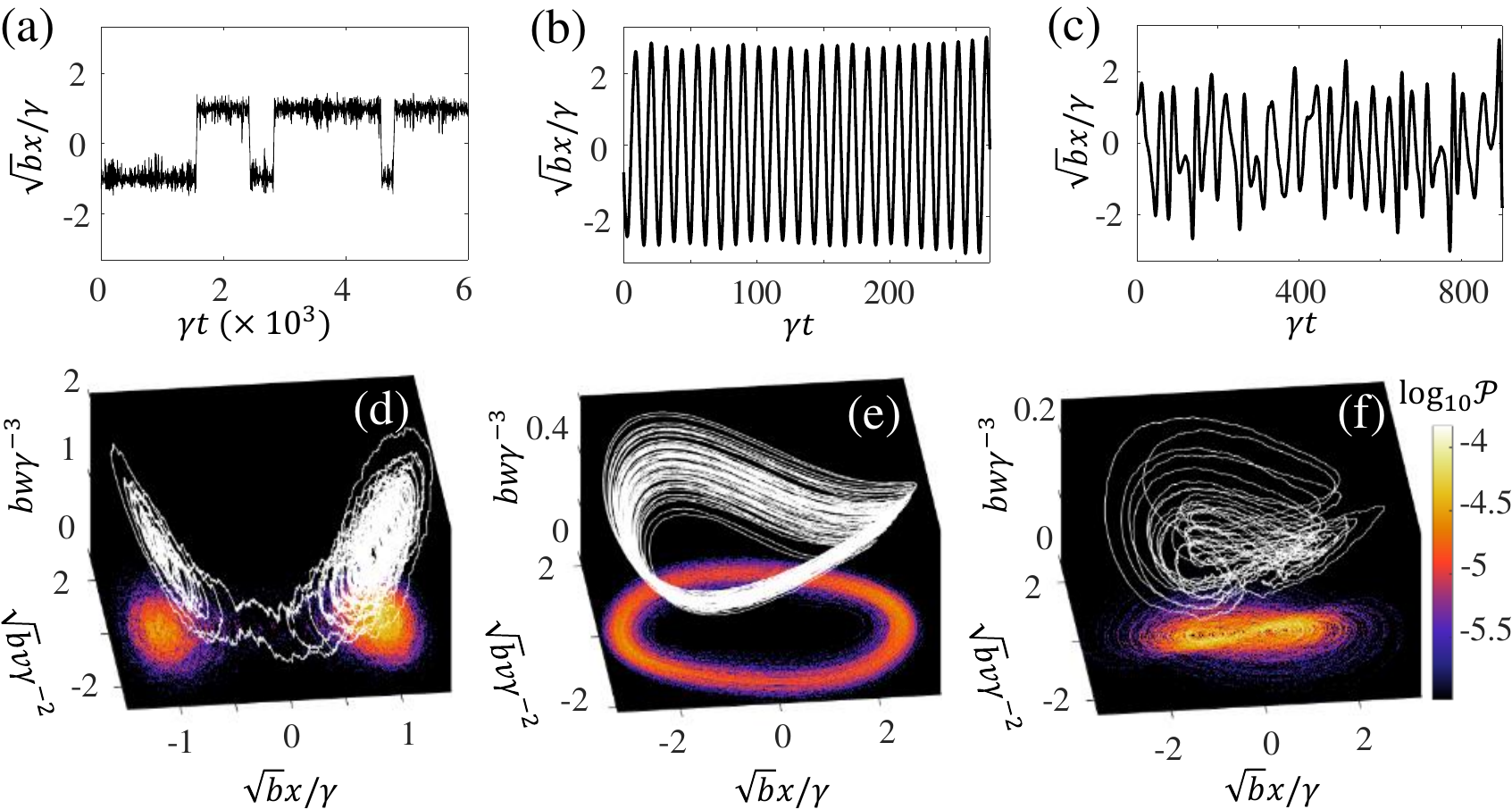}
	\caption{Simulations of an oscillator with Duffing-type nonlinearity having distributed time delay. $\gamma$ is the dissipation rate and $\tau$ is the memory time. (a)-(c) show position as function of time for $\gamma\tau=10^{-2}$, $\gamma\tau=3$, and $\gamma\tau=20$, respectively. (d)-(f) show the phase space trajectories for trajectories of duration $\gamma t=10^3$. The color plot in the $x,v$ plane is a 2D histogram built from a trajectory of duration $\gamma t=10^4$. Simulation parameters: $\gamma=1$, $a=1$, $b=a/10$, $m=10a\gamma^{-2}$, $D^2=\gamma$.}\label{fig:2}
\end{figure}

The dynamics of our nonlinear oscillator strongly depend on the memory time $\tau$. In Fig.~\ref{fig:2}(a) we plot a trajectory of $x$ for $\gamma\tau\ll 1$, i.e., in the limit of an instantaneous nonlinear response. We observe random transitions between the two minima of $V(x)$, located at $x_\pm=\pm\sqrt{a/b}$. This is the typical behavior of a  bistable system. Figure~\ref{fig:2}(d) shows the corresponding trajectory in phase space. The projection of that trajectory on the $x,v$ plane shows the expected behavior for a noise-driven Duffing oscillator without memory. Figures~\ref{fig:2}(b) and ~\ref{fig:2}(e) show a  typical trajectory in time and phase space, respectively, when $\gamma\tau\gtrsim 1$. In that case, we observe stable limit cycles with an amplitude far exceeding the distance between the two minima of $V(x)$. The limit cycles arise due to a Hopf bifurcation near $\gamma\tau = 1$. Finally, for certain ranges of $\gamma\tau>1$, the dynamics become chaotic. An example of this chaotic regime is shown in Figs.~\ref{fig:2}(c,f). Notice in Fig.~\ref{fig:2}(f)  the characteristic shape of the double scroll attractor, indicative of chaos~\cite{Chua86}.

\begin{figure}[!t]
	\includegraphics[width=\textwidth]{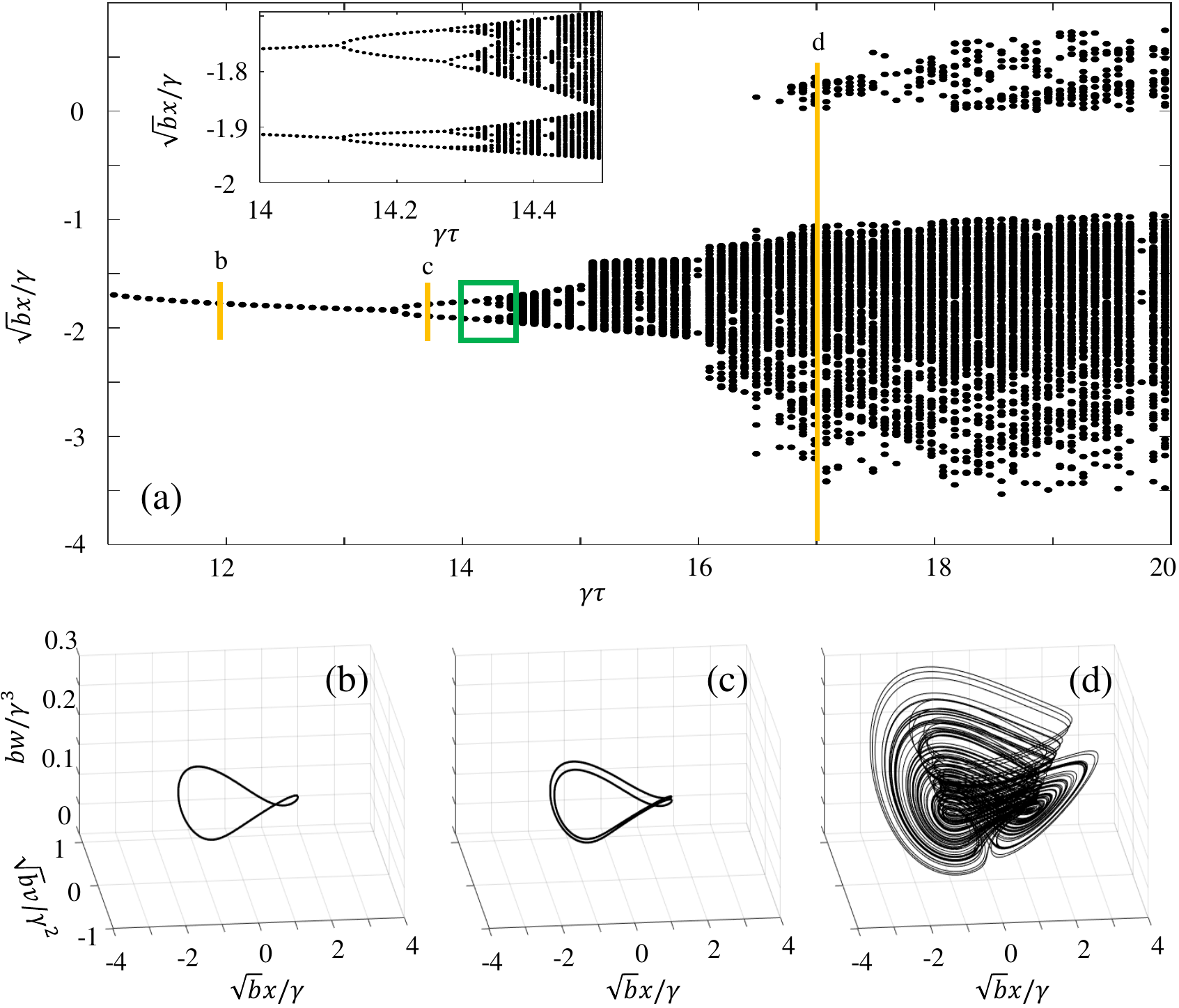}
	\caption{(a) Bifurcation diagram as function of $\gamma\tau$. Vertical yellow lines indicate the phase space trajectories in (b)-(d). Inset: Zoom of region in the green rectangle. Parameters are as in Fig.~\ref{fig:2} with $D=0$.}\label{fig:3}
\end{figure}

Figure~\ref{fig:3}(a) shows a bifurcation diagram for our mechanical oscillator as  $\gamma\tau$ increases. We plot the $x$-values where the phase space trajectory crosses the manifold $v=0$ with $\dot{v}>0$. For $\gamma\tau<13.6$ we observe a single point for each $\gamma\tau$, indicating a period-1 limit cycle [Fig.~\ref{fig:3}(b)]. Near $\gamma\tau=13.6$ we observe a bifurcation, whereafter a period-2 limit cycle arises [Fig.~\ref{fig:3}(c)]. Finally,  for $\gamma\tau>16.5$ we observe the double scroll attractor characteristic of deterministic chaos. The figure as a whole (and the inset in more detail) shows the typical cascade of period-doubling bifurcations leading to chaos. Notice that we also observe periodic windows in between chaotic regimes, occurring within certain ranges of $\gamma\tau$.

\begin{figure}[!t]
	\includegraphics[width=\textwidth]{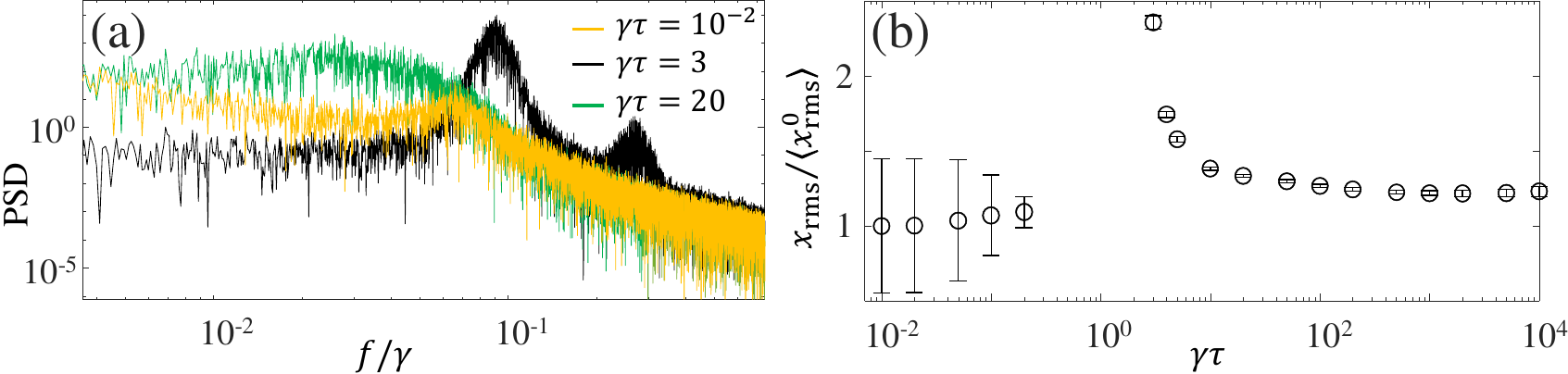}
	\caption{(a) Power spectral density of an oscillator with Duffing-type nonlinearity having distributed time delay. $\gamma$ is the dissipation rate and $\tau$ is the memory time. (b) Enhancement of the root-mean-squared displacement with respect to $\gamma\tau=10^{-2}$ as function of $\gamma\tau$. Parameters are as in Fig.~\ref{fig:2}.}\label{fig:4}
\end{figure}

Figure~\ref{fig:4}(a) shows the spectral response of the oscillator for different $\tau$. The spectra are obtained by Fourier transforming the time traces in Fig.~\ref{fig:2}. For $\gamma\tau=10^{-2}$, the orange curve shows a single shallow peak near the resonance frequency $\omega_\pm$ for the $\tau=0$ case. The peak deviates slightly from $\omega_\pm=a/m$  because of  the finite $\tau$ in our system. Moving on to $\gamma\tau=3$, the black curve in Fig.~\ref{fig:4}(a)  reveals a strong peak at $f/\gamma\approx 0.09$. This peak corresponds to the large amplitude limit cycle oscillations observed in Fig.~\ref{fig:2}(b). We also notice a peak around $f/\gamma\approx 0.27$, due to the oscillations not being purely sinusoidal. Moving on to the chaotic regime ($\gamma\tau=20$), the green curve in Fig.~\ref{fig:4}(a) no longer shows any well-resolved resonances. This is expected  based on the fact that chaotic dynamics can involve a wide range of frequency components. Thinking about nonlinear energy harvesting applications~\cite{cottone2009nonlinear}, one could imagine that chaotic broadband dynamics may be advantageous, since one would like to extract noise power from a wide spectral range. However, Fig.~\ref{fig:4}(a) shows that the power spectrum for the chaotic oscillator decays significantly at high frequencies.  Thus, to assess the frequency-integrated effect of the memory time, let us analyze the root-mean-square (RMS) displacement $x_\mathrm{rms}$ of the oscillator. 

Figure~\ref{fig:4}(b) shows $x_\mathrm{rms}$ normalized to the average value in the small $\tau$ limit, $\left\langle x^0_\mathrm{rms}\right\rangle$, as function of $\gamma\tau$. We observe that $x_\mathrm{rms}/\left\langle x^0_\mathrm{rms}\right\rangle\approx 1$ in the Markovian limit ($\gamma\tau\ll 1$), where memory effects are irrelevant. In contrast, the RMS displacement is greatly enhanced for $\gamma\tau\gtrsim 1$ . As $\gamma\tau$ increases beyond $1$, the RMS displacement decreases because of the chaotic dynamics. For $\gamma\tau\gg 1$ the RMS displacement remains constant as function of $\gamma\tau$, since the system is effectively linear in this regime. However, the RMS displacement is still larger than in the Markovian limit because the system can intermittently make large amplitude excursions and then relax to the monostable state again. Finally, we would like to point out that the lack of data points in the  range $\gamma\tau=0.2-3$ is due to lack of numerical convergence. In our simulations, the amplitude of the limit cycle oscillations diverges in this range. While we believe that more sophisticated numerical methods may resolve this issue, limit cycle oscillations of very large amplitude are still likely to be found in that regime.

To summarize, we have demonstrated how a distributed time delay in a Duffing-type nonlinearity can lead to a rich phenomenology, including the emergence of stable limit cycles and chaos. Remarkably, the amplitude of the limit cycle oscillations can be very large when the delay time is commensurate with the dissipation time. If such a distributed time delay can be realized in nonlinear energy harvesters~\cite{cottone2009nonlinear}, our results could pave the way for massively improving the performance of those systems.

\section*{Acknowledgments}
This work is part of the research programme of the Netherlands Organisation for Scientific Research (NWO). We thank Carlos Pando Lambruschini and Panayotis Panayotaros for organizing the workshop on Advanced Computational and Experimental Techniques in Nonlinear Dynamics, which stimulated this manuscript. We also thank Jason Smith, Aurelien Trichet, and Kiana Malmir for providing the concave mirror used for the experiments in Figure 1. S.R.K.R. acknowledges an ERC Starting Grant with project number 85269. 


\end{document}